\documentclass[aps,pra,twocolumn,groupedaddress,showpacs]{revtex4}
\usepackage{epsfig}
\usepackage{amsmath}
\usepackage{amssymb}
\usepackage{graphicx}
\usepackage{dcolumn}
\usepackage{bm}
\begin{document}

\title{Misconception in  Theory of Quantum Key Distribution\\
-Reply to Renner-
}
\author{Osamu Hirota}
\email{hirota@lab.tamagawa.ac.jp
}

\affiliation{
 Quantum ICT Research Institute,  
Tamagawa University\\
6-1-1, Tamagawa-gakuen, Machida, Tokyo, 194-8610, JAPAN
}


\date{\today}
\begin{abstract}
It has been pointed out  by Yuen that the security theory of 
quantum key distribution(QKD) guided by Shor-Preskill theory has 
 serious defects, in particular their key rate theory is not correct. 
Theory groups of QKD tried to improve several defects.
 Especially, Renner employed trace distance and quantum 
 leftover Hash Lemma. However, the present theory encountered 
a  problem of a quantitative evaluation of security. 
To cope with it, he uses a wrong interpretation on the trace distance and 
its level $\varepsilon_{sec}$, and justifies the unconditional 
security of own system when $\varepsilon_{sec}$ is $10^{-6 } \sim 10^{-20}$. 

In this paper, we discuss the following problems. 
What is the origin of the misconception of the present theory? 
How does the present theory lead to the misconception?. 
To show their process toward the misconception,  
 Koashi-Preskill's theory which has a typical misconception is examined.
 A main point of our comment is that QKD theory ignores  
``the security requirement against attacker" which is necessary 
to compare whole encryption schemes from classical to quantum. 
To clarify it, we emphasize that the trace distance itself cannot 
have any operational meaning such as failure probability, and it is 
only mathematical tool as a measure of closeness. As a result, 
it is given that the security with 
 above values derived from their formulation 
means nothing in the general cryptological sense.
In addition, I point out  that a comment by Bennett and Riedel on 
unconditional security of QKD is not correct.
Also, I point out that the experimental systems of groups of 
Los Alamos, Toshiba-UK, NICT, and others cannot have security 
guarantee even in future.
\end{abstract}
\pacs{03.67.Dd, 42.50.Lc}
\keywords{Quantum key distribution, Shor-Preskill theory, Key rate theory}

\maketitle

\section{Introduction}
The purpose of this paper is to stimulate the real discussion among 
researches who are interested in quantum key distribution (QKD).

So far, taking the recent progress of theory of QKD [1,2] into account,
 I have tried to explain ``What is the main point of misconception on 
the security theory of QKD" in  meetings 
of communities consisting of Information theory, Cryptography, 
Mathematics, and Physics [3]. Recently, Renner has responded [4] to my paper, 
Yuen has responded to it and gave 
the detailed explanation on Renner's fault [5]. 
Here, however, it is useful for the general researchers again 
to explain ``What is a main point of the argument ?", because Renner still 
believes that the system has sufficient security when the level 
of the trace distance $\varepsilon_{sec}$ as the security criterion 
has a order of $\sim 10^{-20}$ [4]. 
His reasoning is that this is very small as 
a probability in physical world. A main subject in this paper is 
to explain that his  $\varepsilon_{sec}$ is not a probability for event.\\

Our claim is as follows:\\

${\bf Claim}$

${\bf (a)}$ The present theory does not deal with the security evaluation 
for QKD correctly. 
At beginning, the correct concept in evaluation of secrecy for 
key distribution had not been employed. That is, they did not care 
``what are they evaluating?". Just it was ``information of Eve" which is 
not an operationally meaningful concept in cryptology.
The information theoretic security in key distribution has to be measured 
by success probability of estimation by Eve according to Shannon [6]
and by her bit error rate (BER) [5] when she makes a wrong estimate of the key.
\\
${\bf (b)}$ The present security theory is originated by Shor-Preskill [7] 
suggesting post processing. Renner introduced a trace distance [8] 
as the security criterion to replace the mutual information criterion 
which is not good quantum mechanically. 
To justify the formulation, he was forced to employ the failure probability 
interpretation of the trace distance( or its level). 
However, it has been already pointed out that such an interpretation 
does not work in view of probability theory. 
That is, the trace distance (or its level) itself cannot have 
interpretation as probability. 
\\
${\bf (c)}$ Why does Renner insist such interpretations in 
the security evaluation ?
The reason is that the present QKD protocol by post processing 
can bound the trace distance or other criteria 
only by very large value in the sense of security, depending on 
phase error estimation and others in BB-84 protocol.
So, in order to justify the quantitative value in their theory, 
they insist on the wrong interpretation of the trace distance  
such as ``failure probability".
By the correct theory of the trace distance, the present theory 
can be re-evaluated.
So the present theory of QKD does not sufficiently ensure the
security and uniformity (against Eve) of generated key sequence by QKD.
\\
${\bf (d)}$ Following our observations on the present 
theory on QKD, the protocol based on post processing cannot 
provide sufficient ``security and uniformity" which is required 
in the common cryptology [3].\\

The research groups of information theory and cryptography in Japan 
already understood our claim by my explanations in several meetings, 
and they also understood that they should take into account the serious 
defect of the security concept in QKD.
However, still the research group of physics ignores the main claim.

To justify our claim, in the following, I will verify concretely 
the misconceptions of the formalism for ``the security" of QKD groups
 in physics by using Koashi's theory as an example of the theory.
To proceed it, I analyze the following technical points.\\
\\
${\bf (1)}$ Theory groups of QKD in physics do not consider the problem
 ``What is a measure of security  in  Shannon's information 
theoretics cryptography ?" 
\\
${\bf (2)}$ The origin of the misconception comes from the fact 
that they do not deal with a concept of security requirement 
against attacker Eve, despite that ``near uniformity of generated key sequence 
against the attacker Eve" has to be ensured. This is related with(1).
\\
${\bf (3)}$ The present theory deals only with privacy amplification to claim 
the uniformity. But, the idea of privacy amplification was introduced 
to create a scale change or scale shift of mutual information between 
the key of Alice and Eve's measurement results in the original papers. 
Although it seems that Shannon's  ${\it mutual}$ ${\it information}$ 
can be reduced by such an operation, it does not mean that the real 
${\it information}$ of Eve is deleted.
Also they ignore the fact that the privacy amplification does not 
improve the success probability of estimation on whole key(or 
uniformity of key sequence against Eve).

\section{Main logic of Security theory of QKD}
\subsection{Basic concept of the present theory}
Here several important sentences claimed by the QKD groups are listed 
as follows [9]:

(1) The task of the quantum key distribution theory in the case of 
finite key is to prove that the key is secure against any wiretapping 
of Eve, up to a small failure probability.

(2) Eve's knowledge can be bounded by the probability that 
she correctly guesses Alice's measurement outcomes. 
This is expressed by the conditional smooth mini-entropy of 
the data from which the key is generated given Eve's quantum system.

(3) The finite case is different from the case of asymptotic rate 
in which perfect security in the limit for $n$ to infinite is considered.

\subsection{Structure of security theory of QKD}
According to Tomamichel et al [9], the final structure of the security theory
to do the above sentences is as follows:\\
In order to define the secrecy of a key, we consider 
\begin{equation}
d=min \frac{1}{2}||\rho_{SE}-\omega_s\otimes \sigma_E||_1\leq \Delta 
\end{equation} 
A key is called $\Delta$-secret from E if it is $\Delta$-close 
to the ideal uniform key.\\

${\bf Definition}$[9]: 
A QKD protocol is called $\varepsilon_{sec}$-secret if it is 
$\varepsilon_{sec}$-indistinguishable from a secret protocol 
where 
\begin{equation}
(1-p_{abort})\Delta \leq \varepsilon_{sec}
\end{equation}

The main misconception is this definition. 
The concept of the secrecy of the protocol is not appropriate to 
evaluate the security of the real cryptological function, 
according to the concept of Shannon's information theoretic security.
This is a kind of play in mathematical school.
However, in order to justify this logic, they provided 
the following structure.[10]

(1) The secrecy of BB-84 follows from the observation that, 
Alice has a choice of encoding a $n$ uniform bits 
in either the X or Z basis, then only one of the following 
two things can be true: either Bob is able to estimate 
Alice's bits accurately if she prepared in the Z basis or
Eve is able to guess Alice's bits accurately if she prepared in the X basis.

(2) Our analysis employs the Quantum Leftover Hash Lemma, which 
gives a direct operational meaning to the smooth mini entropy.
It asserts that, using a random Hash function, it is possible to extract 
a $\Delta$-secret key of length from X, where
\begin{equation}
\Delta =min \frac{1}{2}\sqrt{2^{l-H^{\epsilon'}(X|E')}}+\epsilon'
\end{equation}
$\epsilon'$ is a smoothing parameter.

(3) Following theorem gives a sufficient condition 
for which a protocol $\Phi$ is $\varepsilon_{sec}$-secret. 
The minimum value $\varepsilon_{sec}$ for which it is 
$\varepsilon_{sec}$-secret is called the secrecy of the protocol.\\

${\bf Theorem (R-1)}$[10]: 
 The protocol $\Phi$ is $\varepsilon_{sec}$-secret for some 
$\varepsilon_{sec}>0$ if $l$ satisfies 
\begin{equation}
l\leq max[n(q-h(Q_{tol}+\mu (\epsilon ))-2\log \frac{1}{2\epsilon}
- leack_{EC} -\log \frac{2}{\epsilon_{cor}}]
\end{equation}
where $q,Q,\mu$, and others are given in [10].

\section{Fundamental defect of QKD}
\subsection{Security}
I point out repeatedly here that they have two misconceptions 
as follows:\\

${\bf (a)}$ The information theoretic security can be evaluated 
by the concept of the secrecy of protocol. \\

${\bf (b)}$ There is a reasoning for an interpretation
of failure probability  of the trace distance in their formulation 
which justifies the concept of the secrecy of protocol. \\

The main purpose of this section is to explain why the above two claims are 
misconception ?  To do so, I will make clear first that theory groups of QKD 
misuse the probability theory.

QKD groups assign an interpretation of binary decision to all probability 
measures in their logical process.
A reason of why is that they start from the probability for 
the abort of the protocol in the stage of threshold decision 
based on quantum bit error and they are forced to reach own definition
 of the security: ``${\it the}$ ${\it secrecy}$ ${\it of}$ ${\it protocol}$". 
Thus, they stick to binary events in all logical process.
This is an origin of the misconception as explained below.

Unfortunately, Eq(1) means neither the binary decision nor
the security of protocol. Just it means the closeness.
However, they had started the formulation, keeping the above binary concept. 
After formulation, they realized that they have to include all concepts such as
 ``uniformity",``Eve's guessing probability", and ``indistinguishability" 
which are not binary event.
To proceed it, first, they employ the smooth min entropy which characterizes 
the average probability that Eve guess X correctly. 
Then this is used to give so called Quantum Leftover Hash Lemma 
which discuss a relation between privacy amplification and obtainable key length.
Such a theory may be forcible in connecting Eve's strategy 
and privacy amplification. Finally, they employ the failure probability 
interpretation of the trace distance.
Thus, they claim that the theory is including all 
concepts: ``uniformity",``Eve's guessing probability", 
 and ``indistinguishability", even though the definition of the security is 
binary events such as failure or success.
 
Then numerical examples are given based on this 
formulation, and it is confirmed that one can assume 
 the value of $\varepsilon_{sec} = 10^{-6} \sim 10^{-20}$
 for key length $n \sim 10^4$, because they showed that one can realize 
such values conceptually by  ECC and PAC.
That is, the main protocol based on ECC and PAC can only treat such values 
in the real scheme. But, they cannot handle more small value 
in their formulation, because of the limitation in channel estimation process.
These values are reasonable when $\varepsilon_{sec}$ means indeed 
the failure probability.
Thus they are forced to keep such an interpretation.
\\

Unfortunately, these theories do not deal with the concept of 
the information theoretic security (ITS) correctly, 
which was pioneered by Shannon.
Here I give the answer to the first misconception, introducing 
the correct concept of the ITS by Shannon.\\

${\bf Definition}$: The security of QKD has to be evaluated by 
a success probability:$P_{suc}(K_G)$ of estimation for whole key:$K_G$. 
It is given by quantum $M=2^n$-ary detection theory 
for $n$ qubits sequence. This probability is also a measure of uniformity 
of key sequence.\\

${\bf Theorem(Y-1)}$[5]F Let $|K_G|$ be the length of key.
The success probability of estimation for the whole key can be 
bounded as follows:
\begin{eqnarray}
& & P_{suc}(K_G) \leq  2^{-|K_G|}+ d \\
& & d=min \frac{1}{2}||\rho_{SE}-\omega_s\otimes \sigma_E||_1
\end{eqnarray}
where $d$ is the trace distance.\\

Thus, one can re-evaluate the QKD system (designed by the present theory) 
based on the above relation between the trace distance 
and the correct measure of ITS.

Here one can consider two cases such as
\begin{eqnarray}
A:&&   d \sim 2^{(-|K_G|)},\\
B:&&   d \gg  2^{(-|K_G|)}
\end{eqnarray}
If the system has the case A, it can be regarded as sufficiently secure.
But in real situation, we have the case B. Then we have\\

${\bf Theorem(Y-2)}$[5]FLet $\varepsilon_{sec}$ be 
the average level of $d$ in the Renner's theory.
\begin{equation}
d=min \frac{1}{2}||\rho_{SE}-\omega_s\otimes \sigma_E||_1 \le \varepsilon_{sec}
\end{equation}
Then one has
\begin{equation}
P_{suc}(K_G) \leq  \varepsilon_F
\end{equation}
where $\varepsilon_F=\varepsilon_{sec}^{(1/3)}$ from Markov inequality.\\

Thus, in the case $\varepsilon_{sec}=10^{-20}$ for $|K_G|=10^4$, 
\begin{equation}
P_{suc}(K_G) \leq  10^{-20/3} \sim 10^{-7} 
\end{equation}
If one wants to apply it to one time pad to guarantee $BER=\frac{1}{2}$ 
for any segment of the key, 
the uniformity is required as follows:
\begin{equation}
P_{suc}(K_G) = 2^{-10000} \sim 10^{-3000} 
\end{equation}
\\
Even if one can provide $\varepsilon=10^{-20}$ as individual value
 in Renner's theory, one has $P_{suc}(K_G) \sim 10^{-3000} $ 
vs $P_{suc}(K_G) \sim 10^{-20} $.\\
 Of course, one can relax the uniformity in practice, 
but the difference is so large. Thus one cannot say it is secure.
On the second misconception, I will explain in the next section IV, 
showing the concrete example.

\subsection{Key rate}
The rate theory is formulated as follows:
They believe that the level of the trace distance assumed as
$\varepsilon_{sec}=10^{-6} \sim 10^{-10}$ is good enough already.
So they say that the problem is how many key bits can be extracted under 
this level of the security. The solution is Eq(4) in their formalism, and 
the privacy amplification plays the main role.
Thus one can see that the result of key rate in their discussions
 is effective under the value of $d$(trace distance).
 However, the uniformity of key against Eve is given by her 
$M$-ary quantum detection procedure before the privacy amplification.
The $P_{suc}(K_G)$ cannot be improved by any privacy amplification.
So the uniformity against Eve is fixed before the privacy amplification.
Thus, the relation between the key rate and privacy amplification 
has to be reconsidered. Here one has the following result.\\

${\bf Theorem(Y-3)}$[5]FLet $\varepsilon_{sec}$ be 
the level of $d$ in the Renner's theory and $n$ be the key length. 
Then one has
\begin{equation}
P_{suc}(K_G) \leq  \varepsilon_{F}
\end{equation}
The key rate $R$ under the uniformity guarantee is given by 
\begin{equation}
R \sim \lambda 
\end{equation}
where
\begin{equation}
P_{suc}(K_G) \sim \varepsilon_{F} \equiv 2^{-\lambda n}
\end{equation}
\\
This is completely different from the key rate theory 
based on the present theory. 
Thus, Eq(4) is not able to  provide the sufficient condition for 
extracted key by the theorem(R-1).

\subsection{Asymptotic  vs non-asymptotic key rate}
The present theory has a strange problem such as the title of 
this section. This is because the key rate is discussed by the 
ECC and PAC. However, as one can see in the above section, 
the uniformity is given without ECC and PAC, so  
the problem such as the title of this section has no meaning.
If $\varepsilon_{sec}=10^{-20}$ for the key length $10^4$, 
the key length under the uniformity is about 30 bits and 
the rate becomes zero when bits for the leak from ECC and for
 the authentication are subtracted.

In sum, we emphasize that the present theory is not treating 
the true key rate of QKD. 

\section{Concrete example of process of misconception}
${\bf Koashi}$-${\bf Preskill}$ ${\bf theory}$ \\
Here we describe the structure of Kaoshi-Preskill theory on QKD [11].
 First Koashi explains the concept of security criterion as follows:
 ``The target of the security theory of QKD is to analyze an information 
such as a function of parameters estimated 
from quantum channel" to design  privacy amplification."
 
 In order to proceed such an idea, he provides the following 
mathematical basis.\\
 
 (1)  Let $\rho_1,\rho_2$ be two density operators. In the decision of two 
 density operators, the maximum success probability is given by 
 \begin{equation}
 \frac{1}{2}+\frac{1}{4}||\rho_1 -\rho_2||
 \end{equation}
 From this, one can understand that the trace distance has an operational 
meaning such as ``how much the maximum value of the 
success probability can exceed $\frac{1}{2}$".\\

(2) There is a following relation between fidelity and trace distance.
\begin{equation}
1-F(\rho-\sigma)^{1/2}\leq \frac{1}{2}||\rho-\sigma||\leq 
(1-F(\rho-\sigma))^{1/2}
\end{equation}

(3) Let us assume that there is a little different state from 
the maximum entanglement state, and assume that the fidelity is given by 
\begin{equation}
F(\rho_{AB},\rho_{max})\geq 1-\epsilon 
\end{equation}
The fidelity between state measured by Alice-Bob and ideal state 
$\tau_{ABE}$ is given by
\begin{equation}
F(\sigma _{ABE},\tau_{ABE})\geq 1-\epsilon 
\end{equation}

Koashi says that the fidelity has a strong meaning as follows:\\
Let us use the generated key with the relation of Eq(19) in any system.
But, the probability ``that different situation from the situation 
that the ideal key is used outbreaks" is $\epsilon$. 

Based on such a concept, he proceeds the formulation of the security analysis.
That is, Keys of $n$ bits are described by $M=2^n$ dimensional 
Hilbert space. The density operator for the whole system is given by
\begin{equation}
\sigma_{ABE}=\sum_i\sum_j p_{i,j}|i,j><i,j|_{AB} \otimes \rho_{E}^{(i,j)}
\end{equation}
Here two protocols : X-protocol and Z-protocol are defined as follows:\\

(1) Z-protocol: Alice and Bob determine values of keys 
by classical communication.
The failure probability of the protocol is evaluated by 
\begin{equation}
\eta_Z=1-\frac{1}{M}\sum_i \sum_i p_{i,i}
\end{equation}
(This is average error probability among $M$ events, and it does not mean 
the binary event.)\\

(2) X-protocol: The state of Alice is purified by classical communication.
The ideal state is $|\bar 0>$, and the final state of Alice is $\sigma_A$.
Then the the failure probability is given by
\begin{equation}
\eta_X=1- <\bar 0|\sigma_A|\bar 0>
\end{equation}

Here he employs the trace distance as the measure of difference between 
the real one and the ideal as follows:
\begin{equation}
\eta_{key}=||\rho_{ABE} - \tau_{ABE}^{(key)}||
\end{equation}
By using these concepts, he shows the following theorem.\\

${\bf Theorem(K-1)}[10]$: Let $\eta_X$ and $\eta_Z$ be 
the failure probabilities of
both protocols. The key in the Z-protocol satisfies
\begin{equation}
\eta_{key}\leq 2\eta_Z +2\sqrt{\eta_X}
\end{equation} 
Thus, he showed a relation between the trace distance and 
some $\{\eta \}$ and interpreted it as the failure probability of 
the protocol.
In fact, Tamaki who is a group member of Koashi claims in his paper that 
the protocol will fail just one time at the million trials when the 
level of the trace distance is $10^{-6}$ in that scheme.

Again, one cannot see the reason that each probability measure
 in the process of the security proof means binary such as failure or success.

Lets us mention the problem of the key rate.
In this stage, one needs the error correction code and privacy amplification.
Here values for $\eta_Z$ and $\eta_X$ are assumed, and the security measure
 is fixed by Eq(24). The problem is how many bits can be extracted 
from the shifted key under the Eq(24). 
Koashi says that the privacy amplification reduces 
the correlation between key sequence and Ev's measurement results.
However, this correlation provides only an apparent uniformity 
of key against Eve, because the value of the trace distance is fixed already.  
That is, the basis of the uniformity and the rate is fixed from Theorem (Y-3) 
when the value of the trace distance is fixed.
But, he claims that\\

${\bf Theorem (K-2)}[10]$: When the key sequence after privacy amplification 
is employed as the final key, it satisfies Eq(24), and the secure
 final key length is 
\begin{equation}
l =n-K_Z -K_X
\end{equation}
\\
Thus the logic is the same as that of Renner, and he does not give 
any reasoning of the binary event. Thus it shows clearly
 the process of the misconception from $M$-ary detection towards binary 
 interpretation.

\section{What does the present QKD theory treat ?}
I have clarified the misconception of the present  theory of QKD in the above 
sections. Here I give the general survey of this paper for researchers who are 
not specialist on QKD.\\

(1) ${\bf Criterion}$\\
The theory groups on QKD deal with several criteria such as mutual information 
or trace distance to evaluate the security of own system.
However, these criteria themselves do not have any operational meaning 
of information theoretic security.
So before own formulation of the security analysis, they should consider
 ``what is a security in cryptological sense ?"
But they formulated the upper bound for such criteria without such a 
main discussion. Then, they justified the theory based on binary decision 
such as abort or accept, despite that there does not exist  such a measure
 in their formulation. One can see it from the logic of Koashi 
in the previous section.

According to the original concept on information 
theoretic security by Shannon [6], it is clear that 
a main measure to evaluate a security is the uniformity of key 
against Eve which can be given by a success probability of 
estimation to the key. This has clearly an operational meaning.
Consequently, \\

``${\bf They}$ ${\bf ignored}$ ${\bf the}$ ${\bf Shannon's}$ 
 ${\bf security}$ ${\bf concept}$".\\

As I emphasized in this paper, the relation between such a true 
criterion and trace distance has been given (see Eq(5)).
Thus, one can re-evaluate the quantitative performance of security in 
the present theory, because they discuss the level of the trace distance.
As a result, one can see that the security by the present theory is 
meaningless from Eqs(11),(12). 
 
In addition, the final key rate $R_F$ after subtractions of 
the leak from ECC and message authentication is  as follows:
 \begin{equation}
 R_F \sim 0
 \end{equation}
\\
(2) ${\bf iid}$ ${\bf assumption}$ ${\bf and}$ ${\bf active}$ ${\bf attack}$\\
In the present theory and its application, so many researchers 
keep in mind a serious assumption so called ``iid" in own model, where 
``iid" means independent and identically distributed.
Especially, they believe the effectiveness of privacy amplification 
under such an assumption. However, in the QKD model 
with imperfection, one has to allows Eve to employ an attack 
so called active attack such as bit change or correlation creation. 
So one cannot expect ``iid" in shared key sequence.
Some researchers take it into account, but still their concern is limited to 
effect for privacy amplification only [12].
One should go back to the discussion of the trace distance under 
the active attack, and start under the understanding of 
the fact that the trace distance describes a degree of uniformity 
of the real qubit sequence, not failure probability.

\section{Comment on Experiments}
I here point out that the experimental groups 
of R.Hughes [13],  Z.Yuan and A.J.Shields [14],  M.Sasaki, A.Zeilinger 
and others do not provide the quantitative security evaluation for own systems,
and also do not compare own system with the present theory of the security.
That is, the experiments are independent of the theory, and the progress of the 
theory is ignored. 
Especially A.J.Shields claims [15] that
 ``High bit rate QKD with $100 dB$ security".
Why 1 $Mbit/sec$ is high bit rate nowadays. Also $dB$ means the relative, but 
the security requires absolute value. The technical requirement
 in the real world is at least 1 Gbit/sec $\sim$ 10 Gbit/sec 
if they want to apply the QKD to one time pad encryption.
Even if the present scheme of QKD are implemented, these insecure and
 insufficient systems can be easily replaced by other technology.
 
On the other hand, recently Bennett and Riedel gave a comment
 in March 2013 [16] such that 
``QKD boasts unconditional security even in the presence of 
realistic noise", referring Mayers [17] et al, and ``that the techniques 
have matured enough that small commercial implementations have been explored."
As I have clarified in this paper, there is no proof even for own definition of 
unconditional security. Thus, these comments are not justified. 
I hope that we have an occasion for discussion on this issue each other.

\section{CONCLUSIONS}
I have described the origin of the misconception in the security theory 
of quantum key distribution.
Especially, I pointed out that there are many misconceptions 
in reasonings in the process of the security proof and that 
the security designed by the present theory may be weaker than 
that of the classical cipher, when the system is evaluated correctly.

In order to develop the quantum key distribution and related subjects, 
one should consider more carefully theoretical formulation based on 
quantum detection theory, following ``${\it the}$ ${\it original}$ 
${\it meaning}$ ${\it of}$ 
${\it information}$ ${\it theoretic}$ ${\it security}$ 
${\it of}$ ${\it Shannon}$".
I hope that the fruitful discussion begins by this article.



\section*{Appendix}
\subsection{Discussions in QIT meeting}
[Q-1]: I do not agree with your talk. There are only two results for security 
concept in QKD. That is, it is success or failure of protocol. 
It is enough. (QKD theorist)\\

[A-1]: You should prove it mathematically, because I am showing 
that the trace distance cannot have the meaning of the probability.\\

[Q-2]: Although my field is not QKD, I agree with your comment 
for the mathematical issue  on the trace distance. 
In general, the trace distance is just value as closeness, 
it is not probability indeed. (Theoretical physicist-1)\\

[Q-3]: I think that ``the concept of the security" is vague in any field.  
How to define the concept of security in physics.
(Experimental physicist)\\

[A-2]: QKD is cryptology and not physics. So the researcher of QKD have to 
formulate the security of system as the cryptology, and compare it 
with whole schemes in cryptology. In fact, the definition of 
unconditional security by QKD community has no meaning in cryptology.\\

[Q-4]: What is the difference between unconditional security 
and information theoretic security.(Theoretical physicist-2)\\

[A-3]: The term of ``unconditional security" was invented by QKD community 
such that ``when mutual information of A-E channel or trace distance is bounded
 (sufficiently small) by certain physical parameter, 
the system is called unconditionally secure."
The community thought that this is equivalent to the information theoretic 
security, because it seems to be independent of the computational power.
But this is valid only in the community of QKD. 
That is, it has no operational meaning as the security in the cryptology.\\

[Q-5]: Do you mean that the mutual information (MI) and also trace distance are 
not appropriate criteria. (Theoretical physicist-2)\\

[A-4]: The meaning of the information theoretic security cannot be changed 
in mathematical scheme and physical scheme.

The success probability of estimation for whole shared key by QKD is 
a natural measure of information theoretic security from Shannon's concept.
MI and also trace distance can be used as mathematical tool to bound 
the correct criteria, but these cannot have the operational meaning.
The present theory of QKD is trying to directly give the operational 
meaning to them.
That is, they insist that the trace distance is the failure probability.\\

[Q-6]: What is  the success probability of estimation ?
(Theoretical physicist-3)\\

[A-5]: In ``one time pad" without key distribution, the channel model between 
plaintext sequence $X$ and ciphertext sequence $C$ is described by 
conditional probability (success probability of estimation).
If the channel matrix elements of $X$-$C$ channel are completely uniform 
for all kinds of ciphertext sequences, 
it means the perfect secrecy. That is, $P(X|C)=P(X),\forall C$.
If the $X$-$C$ channel has partial uniformity, it has partial information 
theoretic security. So it provides the quantitative evaluation. 
Here, a main origin of the security comes from the uniformity of pre-shared key 
sequence. 

In the case of QKD, the measurements on key sequence of Eve $Y_E$ on $K_G$ 
suffer the error by quantum effects.
This is also described by the channel model based on success probability 
of estimation $P(K_G|Y_E)$. 
The origin of the security comes from quantum effects of $K_G$-$Y_E$ channel.
Thus, by following Shannon theory, QKD community should employ 
the criterion such as $P(K_G|Y_E)$ based on the detection to each sequence
 in whole possibility :$2^n=M$ at beginning.\\

[Q-7]: You mean when the present QKD is re-evaluated by the correct criterion,
it was extremely bad or weaker than the classical one, 
even if QKD groups claim the unconditional security (by own definition) or 
unbreakable.(Theoretical physicist-3)\\

[A-6]: Yes, it is true. They want to enjoy the terms of intrusion detection 
by uncertainty principle, quantum loophole, and so on, 
but these are parameters only to quantify 
the security criterion. They do not care the real evaluation in cryptology. \\

[Q-8]: It seems that QKD groups do not follow Shannon's basic concept,
and that they want to lock the security issue only in physics. 
Can you analyze why the community of QKD had taken the present way ? 
(Information theorist-1)\\

[A-7]: I am not sure. They thought that QKD is physics.
As you know, QKD is a function in the cryptology. That is, it provides only key 
distribution for symmetric key cipher, though the origin of the basic security 
comes from just physical(quantum) phenomena. As a result, they ignored 
the regular course of the cryptology, as you say.

\end{document}